\documentclass[article, oneside, a4paper]{memoir}
\usepackage[english]{babel}
\usepackage[utf8]{inputenc}
\usepackage[T1]{fontenc}

\usepackage{derivative}  
\usepackage{amsmath, amssymb, amsthm, mathtools, cancel}

\usepackage{graphicx, float, adjustbox, xcolor}
\usepackage[font=small,labelfont=bf,textfont=it]{caption}
\usepackage{hyperref}
    \hypersetup{
        colorlinks,
        linkcolor={red!50!black},
        citecolor={blue!50!black},
        urlcolor={blue!80!black}
    }
\usepackage{cleveref}

\newenvironment{circuit}
  {\crefalias{equation}{circuit}\begin{equation}}
  {\end{equation}\ignorespacesafterend}
\crefname{circuit}{circuit}{circuits}
\Crefname{circuit}{Circuit}{Circuits}

    \newcommand{\half}{\frac{1}{2}} 
    \newcommand{\define}{\equiv} 
    \newcommand{\tensor}{\otimes}

    \newcommand{\argdot}{\makebox[1em]{$\cdot$}}
    \newcommand{\idty}{\mathbb{I}}
    \newcommand{\mat}[1]{\begin{bmatrix*}[r]#1\end{bmatrix*}}

    \newcommand{\qq}[1]{\quad\text{#1}\quad}
    \newcommand{\vb}[1]{\boldsymbol{\mathrm{#1}}}

    \DeclarePairedDelimiter{\expval}{\langle}{\rangle}

    \DeclareMathOperator{\Tr}{Tr}
    \AtBeginDocument{%
      \renewcommand\Re{\operatorname{Re}}
      \renewcommand\Im{\operatorname{Im}}
    }

    \newcommand{\qunaught}{\varnothing}

\NewDocumentCommand{\braket}{ o m m o }{%
    \ensuremath{%
        \IfValueT{#1}{{}_{#1}} 
        \langle #2 | #3 \rangle
        \IfValueT{#4}{_{#4}} 
    }%
}
\NewDocumentCommand{\bra}{ o m }{%
    \ensuremath{%
        \IfValueT{#1}{{}_{#1}} 
        \langle #2 |
    }%
}
\NewDocumentCommand{\ket}{ m o }{%
    \ensuremath{%
        | #1 \rangle
        \IfValueT{#2}{_{#2}} 
    }%
}

\usepackage{tikz}
\usetikzlibrary{quantikz2}  

\newcommand{\qtikzdagger}[1]{\wire[r][1]["\dagger"{above,pos=0.1}, draw=none]{q}}

\usetikzlibrary{calc,arrows.meta}
\newlength{\bswidth}
\newlength{\bsrad}
\tikzset{BSup/.style={
    draw=none,
    minimum width=1.2cm,
    minimum height=1cm,
    path picture={
        \begin{scope}
        \draw[rounded corners=\bsrad] (path picture bounding box.center) +(-0.6, 0.5) -- +(-\bswidth, 0.5) -- +(\bswidth, -0.5) -- +(0.6, -0.5);
        \draw[rounded corners=\bsrad] (path picture bounding box.center) +(-0.6, -0.5) -- +(-\bswidth, -0.5) -- +(\bswidth, 0.5) -- +(0.6, 0.5);
        \filldraw[color=black, fill=white] (path picture bounding box.center) +(-0.4, -0.05) rectangle +(0.4, 0.05);
        \draw[thin, arrows = {-Stealth[inset=1pt, length=3pt, angle'=45, round]}] (path picture bounding box.center) +(0,-0.3) -- +(0,0.3);
        \end{scope}
    },
}}
\tikzset{BSdown/.style={
    draw=none,
    minimum width=1.2cm,
    minimum height=1.2cm,
    path picture={
        \begin{scope}
        \draw[rounded corners=\bsrad] (path picture bounding box.center) +(-0.6, 0.5) -- +(-\bswidth, 0.5) -- +(\bswidth, -0.5) -- +(0.6, -0.5);
        \draw[rounded corners=\bsrad] (path picture bounding box.center) +(-0.6, -0.5) -- +(-\bswidth, -0.5) -- +(\bswidth, 0.5) -- +(0.6, 0.5);
        \filldraw[color=black, fill=white] (path picture bounding box.center) +(-0.4, -0.05) rectangle +(0.4, 0.05);
        \draw[thin, arrows = {Stealth[inset=1pt, length=3pt, angle'=45, round]-}] (path picture bounding box.center) +(0,-0.3) -- +(0,0.3);
        \end{scope}
    },
}}
\newcommand{\bsup}[1]{\gate[2, style={BSup}]{\hspace{5mm}}}  
\newcommand{\bsdown}[1]{\gate[2, style={BSdown}]{\hspace{5mm}}}

\usepackage{csquotes} 
\usepackage[style=numeric-comp, autocite=inline, giveninits=true, sorting=none, url=false, isbn=false, maxbibnames=2]{biblatex}
    \DeclareNameAlias{labelname}{given-family}  
\usepackage{authblk}  
\setlrmarginsandblock{3.5cm}{3.5cm}{*}
\setulmarginsandblock{3cm}{3cm}{*}
\checkandfixthelayout
    
\title{Impact of finite squeezing on near-term quantum computations using GKP qubits}

\author[1,2]{Frederik K. Marqversen}
\author[3]{Andreas B. Michelsen}
\author[2]{Janus H. Wesenberg}
\author[1,2]{Nikolaj T. Zinner}

\affil[1]{Department of Physics and Astronomy, Aarhus University, DK-8000 Aarhus C, Denmark}
\affil[2]{Kvantify ApS, DK-2300 Copenhagen S, Denmark.}
\affil[3]{NNF Quantum Computing Programme, Niels Bohr Institute, University of Copenhagen, Denmark}

\date{July 2025}

\begin{document}

\maketitle
\begin{abstract}
    We present the first detailed simulation of a measurement based quantum computation based on Gottesman-Kitaev-Preskill (GKP) qubits within a quad-rail lattice (QRL) cluster state involving over 100 GKP modes. This was enabled by the recently developed functional matrix product states (FMPS) framework, with which we simulate continuous-variable (CV) quantum circuits while explicitly modelling intrinsic coherent error sources due to finite squeezing. We perform simulated randomized benchmarking across squeezing levels between 5 and 15 dB and find strong agreement with analytical estimates for high quality GKP qubits. As a demonstration of practical computation, we simulate a three-qubit Grover’s algorithm within the QRL and identify a fundamental squeezing threshold—approximately 10 dB—beyond which the algorithm outperforms classical probability bounds.
\end{abstract}

\chapter{Introduction}
This work explores the potential of the Gottesman-Kitaev-Preskill (GKP) encoding for performing practical quantum computation within photonic systems. Theoretically, fault tolerance can be achieved with a source of GKP basis states, Gaussian operations, and with an additional quantum error correction code on the logical level \cite{menicucci_fault-tolerant_2014, noh_fault-tolerant_2020, larsen_fault-tolerant_2021, tzitrin_fault-tolerant_2021}. Previous analyses of fault tolerance have primarily been made under the assumption that high quality GKP states can be modelled by incoherent Gaussian noise applied to ideal infinite energy GKP states \cite{noh_low-overhead_2022, menicucci_fault-tolerant_2014}. To date, GKP qubits have been demonstrated with only very modest amounts of GKP squeezing (< 1 dB) in optical setups \cite{larsen_integrated_2025}, well below the fault-tolerant threshold.
Recent theoretical advances has therefore been aimed at describing the error effects of low-quality GKP states \cite{marqversen_performance_2025,jafarzadeh_logical_2025}. Despite this progress, the computational capabilities of near-term GKP-based systems remain largely unexplored, particularly at the circuit level where the effects of physical GKP states can propagate and interact in non-trivial ways.

A key obstacle has been the lack of efficient classical simulation methods for systems involving many physical GKP qubits. Numerical investigations have previously been limited to systems involving fewer than ten qubits. The introduction of the \emph{functional matrix product states} (FMPS) formalism offers a breakthrough \cite{michelsen_functional_2025}: it significantly outperforms standard tools like Strawberry Fields \cite{killoran_strawberry_2019} for simulating continuous-variable (CV) circuits, and proves especially well-suited for GKP qubits. This advance opens the door to simulations of quantum computations involving significantly larger GKP systems.

While the GKP code supports a fully Gaussian implementation of the discrete-variable (DV) Clifford group \cite{baragiola_all-gaussian_2019}, directly translating DV Clifford gates into CV Gaussian operations is experimentally challenging. Notably, the implementation of key operations such as the phase gate ($P$) and the controlled-$Z$ ($CZ$) gate requires inline squeezing, which is both lossy and difficult to implement at scale. Moreover, the gate-specific nature of these implementations limits flexibility and scalability.

To address these challenges Walshe et al. \cite{walshe_streamlined_2022}, have proposed a measurement-based and error corrected scheme for implementing any Clifford gate using a \emph{quad-rail lattice} (QRL) cluster state (see \cref{fig:qrl}). The QRL is constructed from a stream of so-called “qunaught” states $\ket{\qunaught}$ via time-multiplexing over a low-depth network of beam splitters—a setup already demonstrated experimentally \cite{larsen_deterministic_2021}. Although deterministic qunaught state preparation remains an open problem, the resource generation for the QRL is scalable and experimentally tractable. An additional advantage is that all gates are naturally protected by Knill-type error correction, which has been shown to be optimal for GKP qubits \cite{marqversen_performance_2025}.

In this work, we present the first full physical-level simulations of quantum algorithms using GKP qubits. Our simulations explicitly incorporate the effects of \emph{coherent} noise arising from finite squeezing—noise that is intrinsic to physically realisable GKP states. By excluding external noise sources, we study the fundamental physical limitations imposed by finite squeezing alone. To enable this, we utilise the recently introduced functional matrix product states (FMPS) method for simulating CV quantum systems \cite{michelsen_functional_2025}. Technical details of the simulation methodology are included in the appendices. Leveraging this tool, we simulate QRL-based computations involving over 100 GKP modes, marking a significant advance in the classical study of GKP-based quantum computations.

This paper is structured as follows: In \cref{sec:gkp} we introduce the GKP states that serve as building blocks for the QRL. Then in \cref{sec:rb} we characterize logical error rates via randomized benchmarking \cite{knill_randomized_2008, magesan_robust_2011} across a range of squeezing levels between 5 and 12 dB. Notably, our results show a striking agreement with the predictions of existing analytical error models for high quality GKP qubits, lending strong empirical support to their use—at least within this regime of only intrinsic noise. This validation establishes that within the QRL the theoretical models remain robust even when confronted with the full complexity of coherent physical noise. \Cref{sec:grover} presents a simulation of Grover’s algorithm on the QRL, involving more than 100 physical GKP qubits, as a practical demonstration of the method. The observed algorithmic performance is well predicted by estimated error rates, allowing us to determine the minimum squeezing threshold required to observe genuine quantum behaviour in this setting.

\chapter{Physical GKP states} \label{sec:gkp}
The ideal GKP logical basis states $\ket{0}_L$ and $\ket{1}_L$ are Dirac combs, a superpositions of equidistant quadrature eigenstates of a bosonic mode \cite{gottesman_encoding_2001}:
\begin{gather}
    \ket{0}_L = \sum_{n \in \mathbb{Z}} \ket{2n \sqrt{\pi}}
        ,\quad
    \ket{1}_L = \sum_{n \in \mathbb{Z}} \ket{(2n + 1) \sqrt{\pi}}
        .
    \label{eq:gkp_def}
\end{gather}
where $\ket{q}$ are position eigenstates $\frac{1}{\sqrt{2}}(\hat{a} + \hat{a}^\dagger) \ket{q} = q \ket{q}$. Similarly we introduce a squeezed logical state called the qunaught state
\begin{equation}
    \ket{\qunaught} = \sum_{n \in \mathbb{Z}} \ket{n \sqrt{2 \pi}}.
    \label{eq:qunaught}
\end{equation}
Due to the squeezed spacing, this state is outside of the logical state space; hence the name ``qunaught''. The ideal states in \cref{eq:gkp_def,eq:qunaught} are infinite energy states and cannot be physically realized. Thus physical implementations must use some approximation, of which several have been studied in detail. Here, we consider specifically the photon number dampening approximation. Mathematically the non-unitary photon dampening operator is given by $e^{-\epsilon\hat{N}}$ where $\epsilon \in \mathbb{R}$ and $\hat{N} = \sum_i \hat{n}_i$ is the total photon number operator \cite{menicucci_fault-tolerant_2014}. This particular model has features that we take advantage of in our simulations, the details of which are laid out in \cref{sec:noise model,sec:domain}. Although the choice of approximation is not completely arbitrary we do point to the fact, that the most prominent models including this one has previously been shown to be equivalent \cite{matsuura_equivalence_2020}.

Aligning with the literature we measure the quality of physical GKP states in decibel GKP squeezing. This is the squeezing compared to vacuum of the individual teeth in the GKP state. A photon dampened GKP state $e^{-\epsilon\hat{N}} \ket{\psi}$ with dampening $\epsilon$ has GKP squeezing $s$ given by
\begin{equation}
    s[\epsilon]
        =
    -10 \log_{10}\left( \frac{\tanh(\epsilon/2)}{1/2} \right) \text{ dB}
        \overset{\epsilon \ll 1}{\simeq} 
    -10 \log_{10}(\epsilon) \text{ dB}
\end{equation}
where $1/2$ is the variance of vacuum and $\tanh(\epsilon/2)$ the variance of the teeth of the GKP state \cite{noh_fault-tolerant_2020}.

Given a finite-energy GKP qubit we extract the logical information that it contains as a logical density matrix using a method similar to that laid out in ref. \cite{shaw_logical_2024}. The specifics of this logical decoding is explained in \cref{sec:logical}. We use the logical density matrix $\rho_L$ to define the logical purity $P_L$ of the state simply as the purity of $\rho_L$: $P_L = \Tr(\rho_L^2)$

\chapter{Streamlined Quantum computations with GKP qubits}\label{sec:qrl}
\begin{figure}
    \centering
    \includegraphics[width=0.9\linewidth]{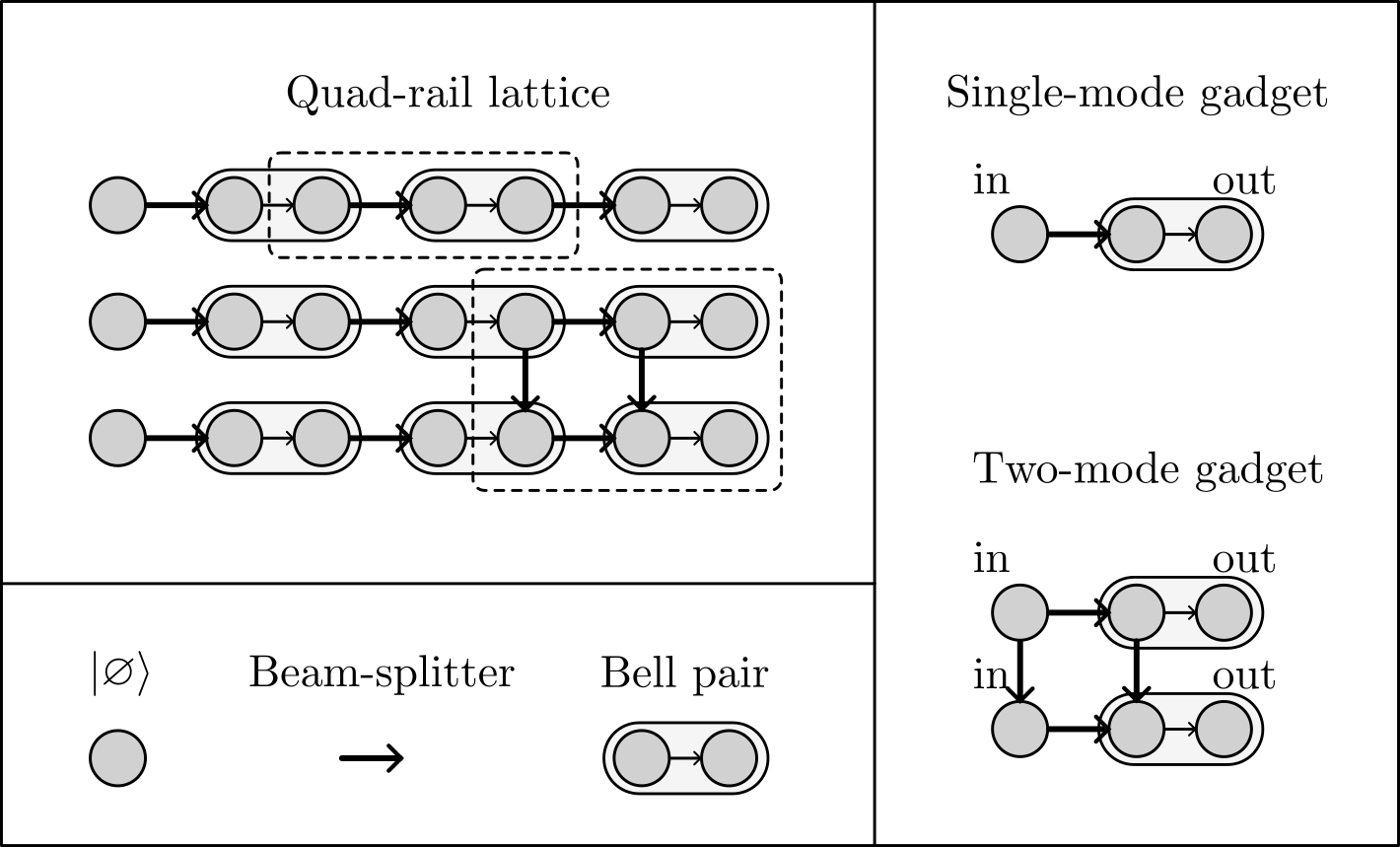}
    \caption{The 2D quad-rail lattice resource state is constructed by concatenating single- and two-mode teleportation gadgets \cite{walshe_streamlined_2022}. A gadget consist of a collection of qunaught states entangled by intersecting on a beam-splitter. A quadrature of each non-output mode is measured by homodyne detection, and by choosing specific quadrature axes deterministic gates between input and output modes are implemented in a measurement-based fashion.}
    \label{fig:qrl}
\end{figure}

In this section, we describe the computational model investigated throughout this work: the \emph{quad-rail lattice} (QRL), a continuous-variable (CV) resource state enabling scalable quantum computation with GKP qubits through measurement-based methods. This model, introduced by Walshe et al. \cite{walshe_continuous-variable_2020, walshe_streamlined_2022}, is built from a network of entangled Bell pairs connected by beam splitters, forming a 2D lattice-like structure (see \cref{fig:qrl}). Although our implementation permits some flexibility in the entanglement structure—enabled by selective control over beam splitters—we retain the term ``QRL'' for consistency.

Computation in the QRL proceeds via homodyne measurements on individual modes, with quadrature angles controlling the applied gates. Single-mode Clifford operations ($\idty$, $H$, $P$, $P^\dagger$) are implemented through \emph{teleportation gadgets} based on Knill-style error correction \cite{knill_scalable_2005, tzitrin_progress_2020, walshe_continuous-variable_2020}, which are embedded directly into the QRL. Two-mode Cliffords, such as \textit{CZ} and \textit{SWAP}, are constructed by linking single-mode gadgets via additional beam splitters. Together, these components enable error-corrected implementation of arbitrary Clifford circuits. Further discussion on the effects and advantages of the Knill-style error correction is provided in \cref{sec:domain,sec:decoding}.

To achieve universality, a non-Clifford gate must be introduced. Walshe et al. propose a heralded method for implementing the $T$ gate \cite{walshe_streamlined_2022}, but to avoid its overhead, we instead employ deterministic magic state injection. Specifically, we replace the standard Bell pair $\ket{\Phi^+}$ in a teleportation gadget with a \emph{magic Bell pair}:
\begin{equation}
    \ket{\Phi^T}
        \define 
    \frac{1}{\sqrt{2}} \left( \ket{00} + e^{i \frac{\pi}{4}} \ket{11} \right)
        =
    T_i \ket{\Phi^+}
\end{equation}
As detailed in \cref{sec:T gate}, these states are prepared offline—using only QRL-native operations—and integrated into the resource wherever a $T$ gate is needed. The method requires measurement-dependent feed forward to adapt homodyne angles, introducing some experimental complexity that could possibly pose a bottleneck for the computations in certain systems.

Pauli gates ($X$, $Y$, $Z$) are handled through \emph{Pauli frame tracking}, a standard strategy that avoids active gate application. This allows correction of measurement-induced Pauli errors without disrupting the lattice by introducing active displacements. Although typically limited to Clifford circuits, we extend this method to allow non-Clifford gates within the QRL framework (\cref{sec:paulis}).

Finally, quantum circuits are compiled into QRL-compatible form by decomposing into native gates and removing explicit Pauli operations. For 2D QRLs, two-mode gates should be localized to adjacent modes to preserve spatial constraints. Once optimized, circuit layers map directly onto the QRL: the number of required Bell pairs scales as $N \times m$, where $N$ is the number of logical qubits and $m$ is the circuit’s layer depth. The total number of GKP qubits used is twice this amount.

\chapter{Error rates in the QRL}\label{sec:rb}

\section{Randomized benchmarking}
Randomized benchmarking is a widely adopted protocol for estimating the average error rates of quantum hardware. Its popularity stems from its broad applicability, robustness to state preparation and measurement errors, and relatively low experimental overhead \cite{knill_randomized_2008, magesan_robust_2011}. The protocol involves applying sequences of randomly selected Clifford gates of varying lengths and measuring the resulting fidelity. These are used to estimate the average fidelity $F(m)$ as a function of circuit depth $m$ of the hardware implementation of Clifford circuits. It turns out that for Clifford circuits $F$ is expected to follow exponential decay as \cite{magesan_robust_2011}
\begin{equation}
    F(m) = A p^m + B
    \label{eq:decay rate}
\end{equation}
where the coefficients $A$ and $B$ capture the effects of state preparation and measurement errors. In particular $B$ equals the $m \to \infty$ limit of the average fidelity over the whole Clifford group. An estimate of the decay parameter $p$ can be then be obtained from the estimated value of $F$ by curve fitting.

The quantity of interest—the average gate error rate $r$—is then inferred from the decay parameter as \cite{magesan_robust_2011}
\begin{equation}
r = (1 - p)(1 - 2^{-N}),
\end{equation}
where $N$ is the number of qubits. The error rate $r$ determined by randomised benchmarking can be associated with single qubit Pauli error probability, or equivalently a depolarising channel with survival rate $1 - \frac{4}{3} r$.

\section{Simulation results} \label{sec:rb results}
\begin{figure}
    \centering
    \includegraphics[width=0.9\linewidth]{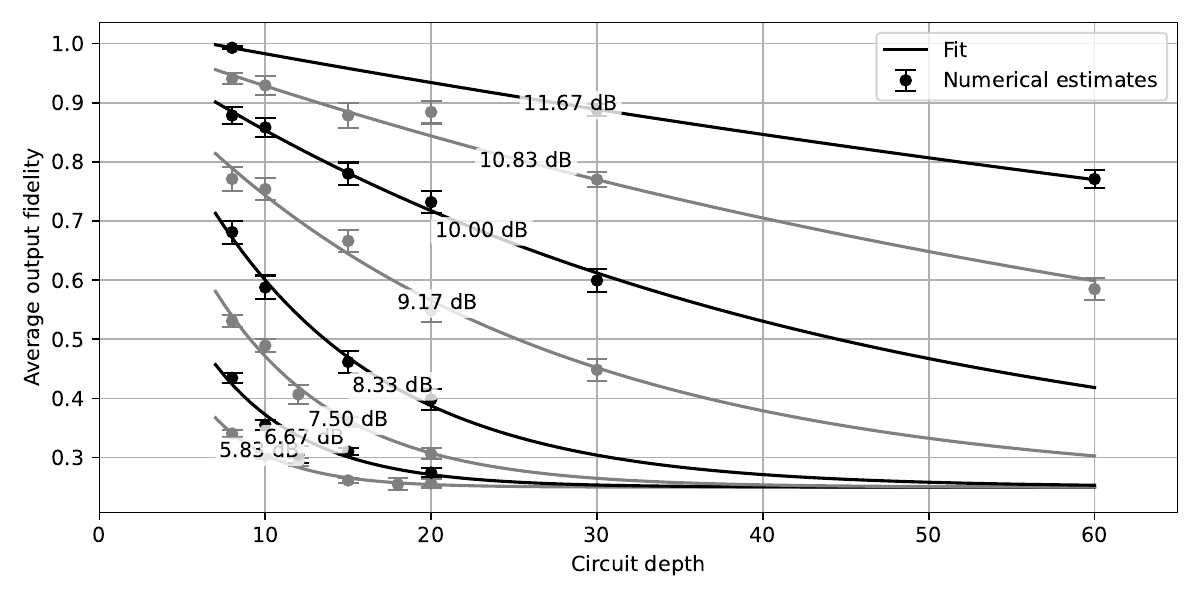}
    \caption{Estimated average fidelity as a function of QRL circuit depth with $1\sigma$ error bars. Lines are fits of the form in \cref{eq:decay rate}. Each fit with corresponding data represents an amounts of GKP squeezing as labelled. The two different shades are only used to distinguish data sets.}
    \label{fig:benchmarking individual}
\end{figure}

\begin{figure}
    \centering
    \includegraphics[width=0.9\linewidth]{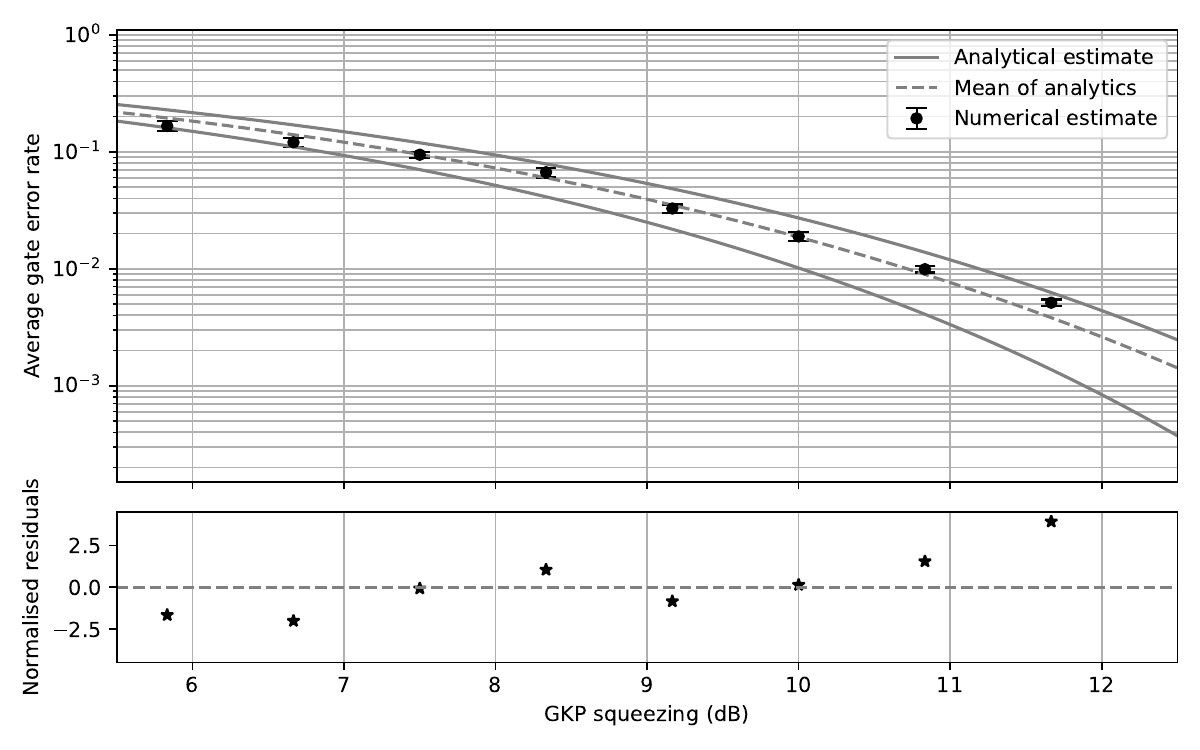}
    \caption{(Top) Numerically estimated average gate error rate $r$ as a function of GKP squeezing with $1\sigma$ error bars. The solid lines are theoretical estimates from ref. \cite{walshe_streamlined_2022} for different gate sets as explained in the main text, and the dashed line is the average of the two solid lines. (Bottom) Normalised residuals between numerical estimates and the dashed mean line.}
    \label{fig:benchmarking results}
\end{figure}

We do randomised benchmarking of computations in the QRL by simulation using the FMPS methods. Specifically we consider $N=2$ qubit random Clifford circuits generated from the set of gates $\{I, H, P, P^\dagger, CZ, SWAP\}$. We exclude the Pauli operators since on the QRL they are implemented in software anyway and thus have no effect on the estimated fidelities. Furthermore, to increase the fitting confidence on the error parameter we explicitly calculate the parameter $B$ for each squeezing level, and find that in all cases it takes exactly the value $1/4 = 2^{-N}$. This is to be expected due to the effects of Clifford twirling \cite{dankert_exact_2009}.
All numerical experiments are shown separately in \cref{fig:benchmarking individual} together with fitted curves of the form in \cref{eq:decay rate}. Results across all experiments are collected in \cref{fig:benchmarking results}. As can be seen from \cref{fig:benchmarking individual}, we include only samples from circuit depths $\geq 7$, since the full Clifford group coverage depth (Cayley graph diameter) of our generating set is 7 (see ref. \cite{marqversen_quantum_computations_2025}).

In \cref{fig:benchmarking results} the two solid lines are the analytical estimates from ref. \cite{walshe_streamlined_2022} that are based on incoherent displacement noise, neglecting the effects of the envelope, and assuming independent $X$ and $Z$ errors. The bottom one is the error rate for identity, Hadamard and swap gates, the top one for phase and controlled $Z$ gates and the dashed line is the arithmetic mean between the two. The two estimates differ because some gates fail due to the intrinsic noise of finite-energy GKP states, whereas some operations also has the effect of amplifying the intrinsic noise leading to slightly elevated error rates. If the analytical estimates are good, it is to be expected that our estimated average gate error rate fall somewhere between the two. As is evident from the figure our benchmarking results match to a high degree the mean. This agreement does seem to fail in the limits of low amounts of squeezing < 7dB and high degrees of squeezing > 11dB. The factors responsible for this behaviour are not entirely clear, but could possibly be the results of interference effects. All in all, this means that for squeezing below 7 dB, the current experimentally attainable level, we can put less faith in the analytical estimates. However, for computationally relevant levels of squeezing above $\sim$10.5db squeezing, corresponding to the surface code fault tolerance limit of about 1\% error rate \cite{fowler_surface_2012}, the larger analytical estimate appears to be an adequate description.

Finally we note that we do not observe any significant decay in logical purity as circuit depth increases. This indicates that the errors that are responsible for the degrading fidelities observed in \cref{fig:benchmarking individual} are not caused by a degrading of the quality of the GKP states, but rather are purely caused by logical errors induced by incorrect decoding of gadget syndromes. This strongly suggests that on the circuit level the noise will be well described by independent rounds of depolarisation noise.


\chapter{Grovers' algorithm in the QRL} \label{sec:grover}

\section{Grovers' algorithm}
Grovers' algorithm is well known as having a provable (quadratic) speed-up compared to the best possible classical counterpart \cite{grover_fast_1996}. The algorithm begins by initialising the system in the uniform superposition state $\ket{+}^{\tensor N} = (H \ket{0})^{\tensor N}$. A sequence of amplitude amplification steps is then applied $r \in O(\sqrt{N/k})$ times where $k$ is the number of valid solutions for the given oracle. We focus on phase-type oracles, meaning that the oracle acts by applying a phase flip $\ket{n} \mapsto -\ket{n}$ to the solutions, and acts as the identity on the rest.

\begin{figure}
    \centering
    \begin{tikzpicture}
        \node[anchor=east, label=below:{\textbf{a)} \hspace{1mm} $\ket{011}$, $\ket{110}$}] at (0, 0) {
            \begin{quantikz}
                & \ctrl{1} &          & \ghost{Z} \\
                & \phase{} & \ctrl{1} & \ghost{Z} \\
                &          & \phase{} & \ghost{Z}
            \end{quantikz}
        };

        \node[label=below:{\textbf{b)} \hspace{1mm} $\ket{000}$, $\ket{100}$}] at (2, 0) {
            \begin{quantikz}
                & \ghost{Z} &          & \\
                & \gate{Z}  & \ctrl{1} & \\
                & \gate{Z}  & \phase{} &
            \end{quantikz}
        };

        \node[anchor=west, label=below:{\textbf{c)} \hspace{1mm} $\ket{010}$, $\ket{111}$}] at (4, 0) {
            \begin{quantikz}
                & \ghost{Z} & \ctrl{1} &          & \\
                & \gate{Z}  & \phase{} & \ctrl{1} & \\
                & \ghost{Z} &          & \phase{} &
            \end{quantikz}
        };
    \end{tikzpicture}
    \caption{Three different phase type oracles and their corresponding solutions.}
    \label{fig:oracles}
\end{figure}
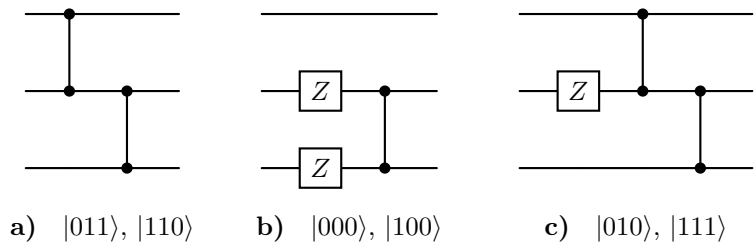

In this work we implement the case $N=3$ and $k=2$ for which the optimal number of iterations is $r=1$. For perfect logical qubits, this instance succeeds with probability 1, evenly distributing probability across the valid solutions. The full circuit is shown in \cref{fig:grover circuit}. In \cref{fig:oracles}, we provide three representative examples of 3-qubit, two-solution, phase-type oracles used in our simulations. A complete list of such oracles is given in ref. \cite{figgatt_complete_2017}.

\begin{figure}
    \centering
    \begin{quantikz}
        \midstick{$\ket{0}$} \gategroup[3, steps=2, style={dashed,rounded corners, inner xsep=2pt}, background, label style={label position=above}]{Initialisation} & \gate{H} & \gate[3]{Oracle} \gategroup[3, steps=7, style={dashed,rounded corners, inner xsep=2pt}, background, label style={label position=above}]{Amplitude amplification} & \gate{H} & \gate{X} & \ctrl{1} & \gate{X} & \gate{H} & \\
        \midstick{$\ket{0}$} & \gate{H} &                  & \gate{H} & \gate{X} & \ctrl{1} & \gate{X} & \gate{H} & \\
        \midstick{$\ket{0}$} & \gate{H} &                  & \gate{H} & \gate{X} & \phase{} & \gate{X} & \gate{H} &
    \end{quantikz}
    \caption{Grovers' algorithm for phase type oracles using three qubits and a single round of amplitude amplification. The circuit generalises trivially to any number of qubits and multiple rounds of amplitude amplification.}
    \label{fig:grover circuit}
\end{figure}
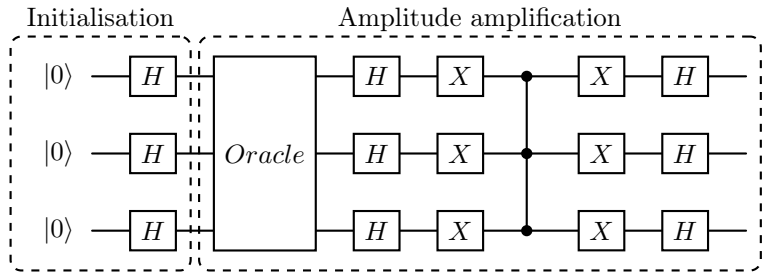

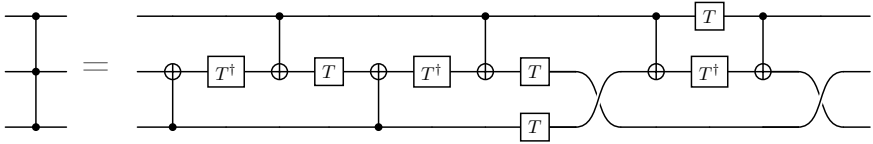
\begin{figure}
    \centering
    \scalebox{0.7}{\begin{quantikz}
        & \ctrl{2} & \ghost{T} \\
        & \ctrl{0} & \ghost{T^\dagger} \\
        & \ctrl{0} & \ghost{T}
    \end{quantikz}}
    {\LARGE = }
    \scalebox{0.7}{\begin{quantikz}
        &           &                   & \ctrl{1}  &           &           &                   & \ctrl{1}  &           &               & \ctrl{1}  & \gate{T}          & \ctrl{1}  &                & \\
        & \targ{}   & \gate{T^\dagger}  & \targ{}   & \gate{T}  & \targ{}   & \gate{T^\dagger}  & \targ{}   & \gate{T}  & \permute{2,1} & \targ{}   & \gate{T^\dagger}  & \targ{}   & \permute{2, 1} & \\
        & \ctrl{-1} &                   &           &           & \ctrl{-1} &                   &           & \gate{T}  &               &           &                   &           &                &
    \end{quantikz}}
    \caption{Optimal Clifford + $T$ decomposition of the $CCZ$ gate \cite[sec. 4.3]{nielsen_quantum_2012}. The $CCZ$ gate is equivalent to the Toffoli gate $CCX$ up to conjugation by Hadamards on the target qubit. Here we have included swap gates in order to obtain a diagram consisting of at only nearest neighbour interactions.}
    \label{fig:CCZ}
\end{figure}

To execute the circuit on the QRL, it must be transpiled into its native gate set. Clifford gates are naturally supported via teleportation gadgets, making them efficient to implement. However, non-Clifford operations—like the $CCZ$ gate in the circuit in \cref{fig:grover circuit}—must be decomposed into $T$ gates and Clifford gates. The decomposition used here is optimized for QRL execution using only nearest-neighbour interactions, as shown in \cref{fig:CCZ}, at the expense of two additional SWAP gates.

\section{Simulation results}
We simulate the three-qubit Grovers' algorithm in the QRL using each of the three representative oracles from \cref{fig:oracles}. The circuit in full has QRL depth of 18 (17 in the case of the $\ket{000}$ and $\ket{100}$ oracle) and width 3. This gives a total of 54 GKP Bell pairs, 7 of which are magic Bell states. In terms of the size of the simulation this corresponds to a total of $54 \times 2 = 108$ separate physical GKP modes.

\begin{figure}
    \centering
    \includegraphics[width=0.9\textwidth]{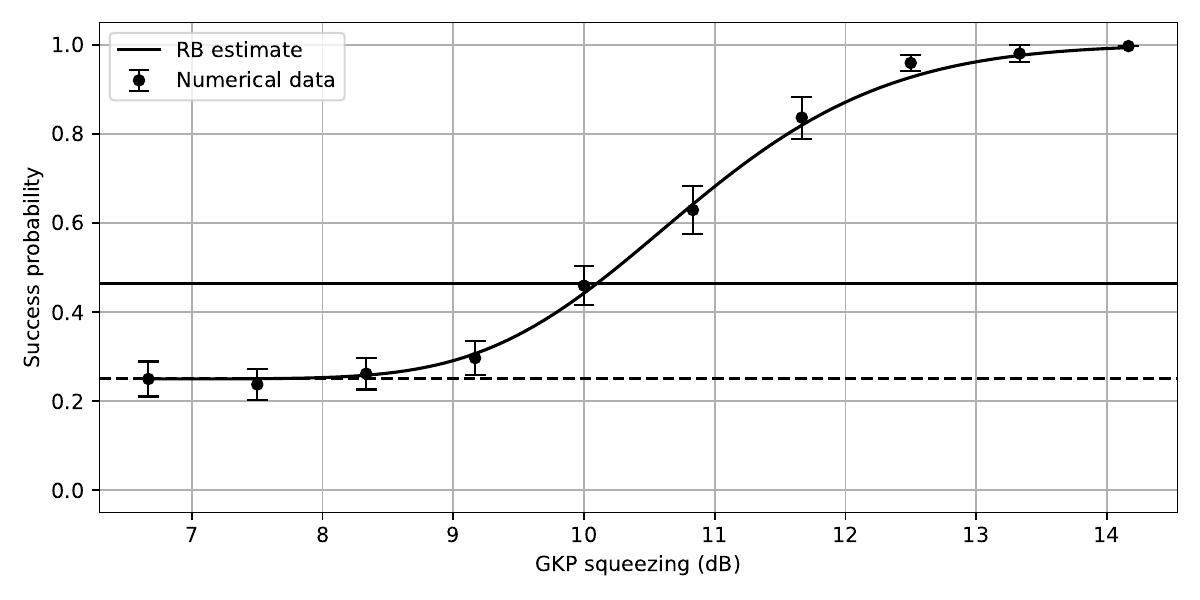}
    \caption{The probability of obtaining a correct solution from a single run of Grovers' algorithm as a function of GKP squeezing. Error bars indicate 95\% confidence intervals. The analytical estimate is obtained by assuming depolarisation noise with strength as determined by the randomised benchmarking. The dashed horizontal line at $2/8$ corresponds to a random output with all logical outputs equally likely. The solid horizontal line at $13/28$ marks the probability of success for a classical search when only one query to the oracle is permitted.}
    \label{fig:grover success}
\end{figure}

\begin{figure}
    \centering
    \includegraphics[width=0.9\textwidth]{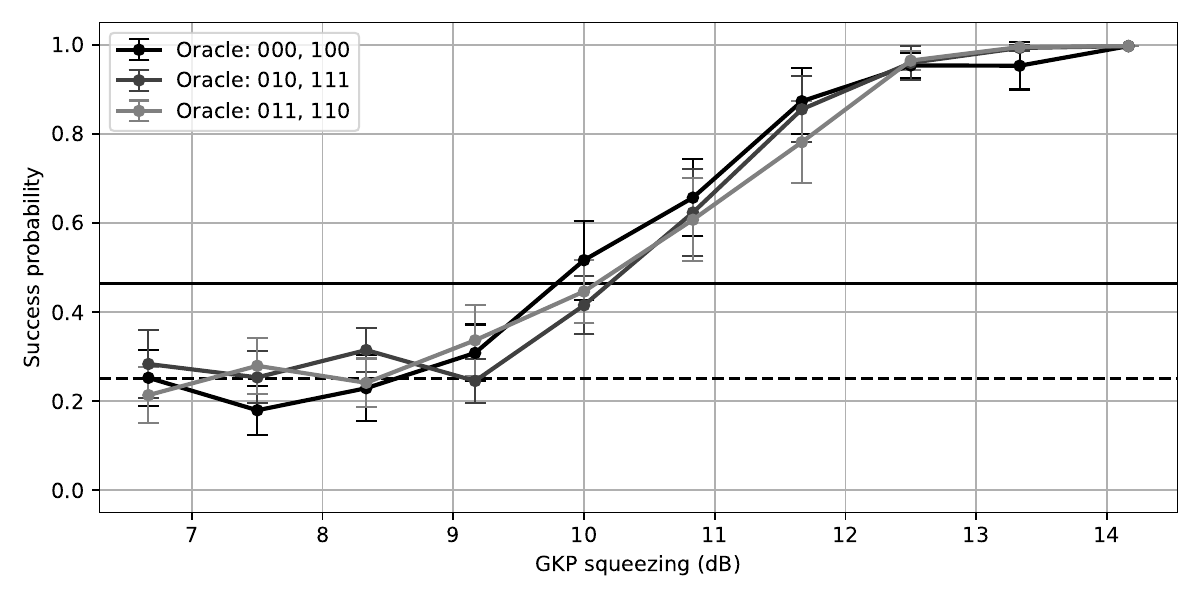}
    \caption{The probability of obtaining a correct solution from a single run of Grovers' algorithm with each of three different oracles. Error bars show $2\sigma$ confidence intervals.}
    \label{fig:grover success oracles}
\end{figure}

Our numerical findings are presented in \cref{fig:grover success,fig:grover success oracles}. We observe no significant difference between different oracles (see \cref{fig:grover success oracles}), which allows us to make general observations on the performance independent of oracle. For this we collect all samples across oracles in \cref{fig:grover success}. In \cref{fig:grover success} we show the probability that a single run of the algorithm correctly outputs any of the two states tagged by the given oracle as a function of GKP squeezing. This probability is the combination of the probability of the algorithm outputting a given state and the probability of obtaining a correct result when doing the final $Z$ basis readout of that output state. Probabilities are estimated from sampling of the circuit across all values of GKP squeezing.

From our simulations we observe that for GKP squeezing above $\sim 10$ dB the probability of obtaining a solution is greater than the classical counterpart. For oracles of this size, such a result is obviously not of any practical interest. It does however provide a natural proof of quantumness. We emphasize again that our results apply to systems with no external noise. So what we can conclude is that $\sim 10$ dB of GKP squeezing is a fundamental lower bound on the amount of squeezing needed for a physical system running this algorithm to exhibit quantumness. Any less and the intrinsic errors fundamentally tied to any real physical system drown out the advantage of the algorithm.

In \cref{fig:grover success} we also include an analytical estimate. This estimate is obtained by assuming single qubit Pauli errors with probability $r$ given by the analytical average shown as a dashed line in \cref{fig:benchmarking results}. This corresponds to a depolarisation channel with survival rate $p = 1 - \frac{4}{3} r$. The survival rate of a circuit is then approximately $p^{N d}$ where $d$ is the circuit depth and $N$ the number of qubits or width of the circuit. With probability $1 - p^{N d}$ the output will be a maximally mixed state:
\begin{equation}
    \rho_\text{out} = p^{N d} \rho_\text{true} + (1 - p^{N d}) \idty / 2^N
\end{equation}
where $\rho_{true}$ is the output of the noiseless circuit. The probability $p_\text{success}$ of successfully obtaining one of the $k$ solutions is then readily obtained from this. 
\begin{equation}
    p_\text{success} \simeq p^{N d} p_\text{success}^\text{true} + (1 - p^{N d}) \frac{k}{2^N}
        \qq{with}
    p = 1 - \frac{4}{3} r
\end{equation}
Estimating the probability $p_\text{success}^\text{true}$ is done by standard analysis of Grovers' algorithm. In our specific case with $N=3$ and $k=2$ the noiseless circuit produce a clean superposition of the solution states resulting in $p_\text{success}^\text{true} = 1$. This estimate is shown as a solid line in \cref{fig:grover success}. From the figure it is evident that this model captures the general performance quite well.

\chapter{Conclusion}
In this work, we have demonstrated large-scale simulations of GKP-based quantum computation using the FMPS method. Our simulations focus exclusively on coherent noise from finite GKP squeezing, offering a clear window into the intrinsic limitations of physical implementations.

Randomized benchmarking shows a high degree of agreement with existing analytical estimates for the gate error rates. This supports the validity of using these models even when considering coherent error propagation across deep circuits. Our results suggest that logical-level depolarizing noise provides a good approximation of the physical error mechanisms, as evidenced by consistent error rate scaling and the absence of logical purity decay with circuit depth.

Crucially, we simulate a three-qubit Grover’s algorithm and show that algorithmic success probabilities exceed classical bounds only when the squeezing exceeds approximately 10 dB. This establishes a fundamental lower bound on the physical squeezing required for a physical implementation to exhibit detectable quantum behaviour. The ability to simulate this threshold accurately both confirms the utility of the FMPS method and also allow us to review the physical hardware requirements needed for near term computations using GKP qubits.

Our simulation methods represent a significant advance in the classical tools available for studying continuous-variable quantum computing. These tools form a crucial part of the roadmap toward experimental efforts aimed at realising practical quantum computing with GKP qubits.

\chapter{Acknowledgements}
This work was supported by Innovation Fund Denmark under grant no. 1063-00046B - “PhotoQ Photonic Quantum Computing” and by the Novo Nordisk Foundation, Grant number NNF22SA0081175, NNF Quantum Computing Programme.

We would like to thank Ulrik Lund Andersen and Michael Kastoryano for many fruitful discussions and collaborations on related projects, and we also thank our colleagues Anton Alnor, Josefine Robl, Sebastian Yde Madsen, and Ulrich Busk Hoff, as well as the rest of the research team at Kvantify.

\chapter{Code availability}
The simulation software used in this work is optimised and built for simulations of arbitrary circuits in the QRL and is available at ref. \cite{marqversen_quantum_computations_2025}.

\emergencystretch=1em
\printbibliography

\appendix
\chapter{Universal streamlined quantum computation in the QRL}\label{app:qrl details}
\section{GKP error decoding}\label{sec:decoding}
The action of the teleportation gadgets is only deterministic up to a known phase space displacement that depends on the measurement outcomes. From ref. \cite{walshe_continuous-variable_2020} we know that specifically, the single-mode gadget adds the following displacement:
\begin{equation}
    \vb{s} = \mat{s_1 \\ s_2} = \mat{\sqrt{2} \Re(\mu_{a,b}) \\ \sqrt{2} \Im(\mu_{a,b})}
        \qq{with}
    \mu_{a,b} = i\frac{m_a e^{i\theta_b} + m_b e^{i\theta_a}}{\sin(\theta_a - \theta_b)}
        \label{eq:displacement error}
\end{equation}
where quadrature angles are measured from the $q$-quadrature ($\theta=0$ is a $q$-quadrature measurement). The two-mode gadget, on the other hand, introduces two-mode displacement as described in ref. \cite{walshe_streamlined_2022}. However, decoding is done simply by decoding the displacement of each mode independently using the same decoder as for the single-mode gadget. 

Decoding is the problem of determining what Pauli error was introduced by this displacement. Here we use the parity of the closest integer multiple of $\sqrt{\pi}$ in each quadrature of the measured displacement $\vb{s}$ defined in \cref{eq:displacement error}:
\begin{equation}
    X^{n(s_1)} Z^{n(s_2)}
        \qq{with}
    n(x) = [x/\sqrt{\pi}] \text{ mod } 2 \in \{0, 1\}
\end{equation}
where $[\argdot]$ denotes rounding to nearest integer. The numbers $n(s_1)$ and $n(s_2)$ are referred to as the $X$ and $Z$ GKP syndromes.

Generally we want to minimise the amplitude (average displacement) of states during computation, in order to minimise photon losses. Since displacements of the GKP code correspond to Pauli operations, this is often taken care of as part of the GKP decoding. However, Knill type error correction with qunaught states has the feature that the teeth of the error corrected state inherit their positions directly from the ancilla Bell states that are used \cite{marqversen_performance_2025}. Since the Bell states are independently prepared states, these will in general already have minimal amplitude, implying that produced error corrected states will as well. In summary, the logical displacement correction should itself be chosen minimal, which is why we include a $(\text{mod } 2)$ operation in our decoder.

Furthermore, this particular feature implies that the measured syndromes from one round of error correction to the next are independent, showing that there is no advantage to gain from some correlated decoding scheme. An additional property of Knill type error correction with qunaught states is that, since the Bell states have symmetric envelope between the two modes, the error corrected state will have symmetric noise between the two quadratures \cite{marqversen_performance_2025}. Thus, we can expect $X$ and $Z$ errors to be equally likely when computing in the QRL.

\section{T gate in the QRL}\label{sec:T gate}
In order to achieve a universal gate set we must add at least one non-Clifford gate. For the GKP code this is usually the $T$ gate implemented by magic state injection \cite{gottesman_encoding_2001, tzitrin_progress_2020, baragiola_all-gaussian_2019}. A particularly convenient feature of the GKP code is that the magic T-state $\ket{T} = T \ket{+}$ is experimentally simple to obtain, since it is distillable by performing error correction of the vacuum state \cite{baragiola_all-gaussian_2019}. The original work of B. W. Walshe et al. \cite{walshe_continuous-variable_2020, walshe_streamlined_2022} suggests a heralded approach which fits seamlessly within the QRL. In order to avoid the large repetition overhead such a method entails, we propose an alternative deterministic method by injection of externally prepared magic states.

First we define the magic Bell state
\begin{equation}
    \ket{\Phi^T}
        \define 
    \frac{1}{\sqrt{2}} \left( \ket{00} + e^{i \frac{\pi}{4}} \ket{11} \right)
        =
    T_j \ket{\Phi^+}
\end{equation}
where $j$ can be either of the two modes. Using the results and methods of ref. \cite{marqversen_performance_2025} one can confirm that by using the magic Bell state in the single-mode gadget one gets
\begin{circuit}
\begin{quantikz}[row sep={1cm, between origins}]
    \lstick{$\ket{\psi}$}      & \bsup{}   & \meterD{\hat{q}}  & \setwiretype{c} \rstick{$m_a$} \\
    \lstick[2]{$\ket{\Phi^T}$} &           & \meterD{\hat{p}}  & \setwiretype{c} \rstick{$m_b$} \\
                               & \ghost{R} & \rstick{$T A(\qunaught, \qunaught) D(\mu_{a,b}) \ket{\psi}$}
\end{quantikz}
\end{circuit}
where we have used the notation from ref. \cite{walshe_continuous-variable_2020}, and where $A(\qunaught, \qunaught)$ is the approximate GKP code projector. The only difference is the addition of a $T$ gate to the output.

As discussed in \cref{sec:decoding}, the introduced displacement operator introduces a logical Pauli error. In this case, the error sits in front of the $T$ gate. For the sake of argument, let the Pauli error be $E$. In order to transfer the $T$ gate to the state we must commute the two:
\begin{equation}
    T E = E (E^\dagger T E) = E (P^\dagger)^{\delta_{X \in E}} T
\end{equation}
where $\delta_{X \in E} = 1$ if $X \in E$ and $= 0$ otherwise. We find that in order to apply a $T$ gate deterministically, we have to, conditioned on $X \in E$, correct the state by applying a $P$ gate to cancel the introduced $P^\dagger$. This exactly mirrors the effects of DV magic state injection. Additionally we note that since $P^\dagger T = T^\dagger$ this also shows that a $T^\dagger$ gate can be implemented by using the same magic Bell pair simply by using the negated condition instead. This detail is important for the seamless implementation of $T$ gates in the QRL as will be discussed in \cref{sec:paulis}.

The conditional $P$ gate can be seamlessly incorporated into the QRL in the following way: Make space for an additional single-mode gadget immediately following any $T$ gate. If the Pauli error of the $T$ gate includes the $X$ operator, then the homodyne measurements of the gadget are chosen to implement a $P$ gate. If not, then the measurements are chosen as to implement an identity $I$.

Finally we note that the magic Bell state can be prepared by the following circuit
\begin{circuit}
\begin{quantikz}[row sep={1cm, between origins}]
    \lstick{$\ket{T}$} & \ctrl{1} & \rstick[2]{$\ket{\Phi^T}$} \\
    \lstick{$\ket{0}$}   & \targ{}  &
\end{quantikz}
\end{circuit}
A magic Bell pair factory could then be designed as a QRL computation as follows: First a magic state $\ket{T} = T \ket{+}$ is distilled from vacuum by applying a sequence of error corrected identity gates within a QRL \cite{baragiola_all-gaussian_2019}. Upon success, the above circuit is applied deterministically using a single two-mode gadget.

\section{Pauli operators} \label{sec:paulis}
Although the Pauli operators can be obtained from the gate set available using the single-mode gadget, it would be advantageous to have these at our disposal as well. In particular because, as discussed in \cref{sec:paulis}, in order to apply gates deterministically within the QRL, we must be able to correct the introduced logical syndrome errors following each gadget. That is, we must be able to apply conditional Pauli operators. 

There is no direct way of seamlessly integrating Pauli operations into the QRL due to the fact that Pauli operators for the GKP code are displacement operators which are active operations. However, by taking inspiration from quantum error correction, it turns out that one can actually entirely circumvent ever having to apply Pauli operators explicitly during a computation simply by keeping track of the accumulated Pauli operator in software. We start by briefly explaining how this is done in the well known case of Clifford gates before discussing non-Cliffords.

Let $\ket{\psi}$ be the the expected state produced by some DV logical circuit. The result of running this circuit in the QRL will be $P \ket{\psi}$ where $P$ is some accumulated Pauli operator that has yet to be applied. Acting with Clifford operator $C$ produce
\begin{equation}
    C P \ket{\psi} = (C P C^\dagger) C \ket{\psi} = P' C \ket{\psi}
\end{equation}
where $P'$ is a new accumulated Pauli operator since $C$ is Clifford. So the result of applying $C$ to the non-corrected state is the same as applying $C$ to the corrected state iff we update the stored accumulated Pauli operator.

Clearly this procedure will not work in the case of non-Clifford gates. However, we have only one non-Clifford gate to consider, namely the $T$ gate. The solution to Pauli tracking with $T$ gates is to commute the Pauli operator through the $T$ gate. Notice how the situation changes
\begin{equation}
    T_i P \ket{\psi} = P (P^\dagger T_i P) \ket{\psi} = P T_i^{\dagger_{X_i \in P}} \ket{\psi}
\end{equation}
where $\dagger_{X_i \in P} = \dagger$ if $X_i \in P$ and $= 1$ otherwise. This follows from the fact that for the $T$ gate we have $X T X = T^\dagger$. So, if the accumulated Pauli does not include an $X_i$ correction $X_i \notin P$, then the effect of applying a $T$ gate to the uncorrected state is the same as applying a $T$ gate to the corrected state with the same Pauli correction. However, if on the other hand $X_i \in P$, then the effect on the corrected state is that of a $T^\dagger$. The relation is easily reformulated as
\begin{equation}
    T_i^{\dagger_{X_i \in P}} P \ket{\psi} = P T_i \ket{\psi}
\end{equation}
So when we do Pauli tracking and encounter a $T$ gate, we simply consider the currently accumulated Pauli correction $P$, and if $X_i \in P$ we do not apply a $T$ gate but rather a $T^\dagger$. In this way, we still obtain the desired action of applying a $T$ gate to the corrected state without having to explicitly correct the Pauli correction.

An important note is that this conditional $T$/$T^\dagger$ gate actually fits seamlessly into the QRL due to the fact that the two gates are realised using the same magic Bell state as discussed in \cref{sec:T gate}. This means that the QRL resource can be prepared to apply $T$ gates deterministically even without a priori knowledge of the Pauli correction at that point. We have thus shown that Pauli tracking can be applied to any circuit in the QRL, and so, Pauli operators i.e. displacement operators never have to explicitly be implemented.

\chapter{Details on simulations of QRLs using FMPSs} \label{app:simulation details}
In order to simulate the large number of GKP qubits required to run a circuit on the QRL we use the recently developed methods of FMPS for simulating CV quantum computations \cite{michelsen_functional_2025}. For the purpose of this work, these methods are implemented for directly simulating the single- and two-mode gadgets specifically. In this section we review details that explicitly relate to this present work, and that were not discussed in the original.

\section{Randomised singular value decomposition}
As discussed in ref. \cite{michelsen_functional_2025}, the effectiveness of the FMPS simulations rely on retaining small internal dimensions of the matrix product states, done in a way that keeps all of the most important information without losing too much of the details. This is done is by performing truncated singular value decomposition (SVD), in which all small singular values are discarded. 

The main contributor to computation time is performing truncated SVDs. And so, the effectiveness of the FMPS simulations rely on an effective algorithm for performing SVD. The parameters needed to obtain a faithful CV representation induce some extremely demanding cases in terms of SVD. So much so, that exact SVD is not feasible, sometimes in terms of time, others even in terms of space/memory.

In order to overcome this, the simulations presented in this work have been performed using an approximate truncated SVD algorithm, implemented by employing the Randomized Range Finding algorithm introduced in ref. \cite{halko_finding_2010}. The algorithm relies on the fact that the given matrix has low rank. In practice, we observe that the internal dimensions are capped somewhere on the order 64. Intuitively, we can convince ourselves that this will always be the case due to the fact that GKP states approximates DV qubit states, and so, we expect that the stored amount of information is actually mainly of a DV nature, which in general is much less than that of CV information.

\section{Noise model}\label{sec:noise model}
The noise model that we consider in this work is described by the non-unitary photon dampening operator given by $e^{-\epsilon\hat{N}}$ where $\epsilon \in \mathbb{R}$ and $\hat{N} = \sum_i \hat{n}_i$ is the total photon number operator \cite{menicucci_fault-tolerant_2014}. The GKP basis state wave functions can be computed exactly and the results can be found in ref. \cite{matsuura_equivalence_2020}. This noise model has the particular feature that it produces states that have spherically symmetric envelopes in phase space. To see this, let $U$ be a general phase space rotation also called a passive linear transformation, meaning that it is a combination of beam-splitters and phase rotations. Then $U$ preserves total photon count $U \hat{N} U^\dagger = \hat{N}$ and so, commutes with the photon dampening operator
\begin{equation}
    U e^{-\epsilon\hat{N}} \ket{\psi} = e^{-\epsilon\hat{N}} U \ket{\psi}
\end{equation}
If $\ket{\psi}$ is an ideal GKP qubit, it will extend infinitely in all directions. This will obviously be true for any rotation $U \ket{\psi}$ as well. Thus the two states will have the same envelope when acted on by the photon number dampening operator. And so, we conclude that the envelope of GKP qubits under the photon dampening approximation are symmetric under any phase space rotations $U$. In other words, the envelope is spherically symmetric.

\section{Logical decoding of states}\label{sec:logical}
The logical information of a GKP state can be extracted into a logical density matrix by the method presented in ref. \cite[appendix D]{shaw_logical_2024}. Our presentation here is only for single mode states, however, the idea generalises trivially. The DV Pauli operators forms a basis for the density matrices, and in particular we get the following decomposition of single qubit density matrices:
\begin{equation}
    \rho = \idty \Tr(\rho) + X \Tr(\rho X) + Y \Tr(\rho Y) + Z \Tr(\rho Z)
\end{equation}
Since $\Tr(\rho \sigma) = \expval{\sigma}$ is nothing but an expectation value, we change out the expectation value of DV Pauli operator $\sigma$ over a DV qubit to that of a CV GKP Pauli operator $\sigma_{CV}$ and CV GKP state $\rho_{CV}$. Thus
\begin{equation}
    \rho_L = \idty \Tr(\rho_{CV}) + X \Tr(\rho_{CV} X_{CV}) + Y \Tr(\rho_{CV} Y_{CV}) + Z \Tr(\rho_{CV} Z_{CV})
\end{equation}
The GKP Pauli operators are displacement operators which are easy to simulate on a FMPS. And so, the CV expectation values in the above expression can be efficiently determined from a FMPS.

Given any CV state the corresponding logical density matrix directly encodes the measurement probabilities of any GKP Pauli measurement. In particular, for a state with logical density matrix $\rho_L$, the probability $p$ of measuring a CV outcome consistent with logical DV state $\ket{n}$ is simply given by $p = \bra{n} \rho_L \ket{n}$.

\section{Optimal gadget connectivity}
Retaining FMPS structure is fundamental to the efficiency of the simulation method. The main time cost of the simulation as a whole is performing the SVD's that follow any multi-mode gate required to restore the FMPS structure. For a two-mode gate, the number of SVDs required is equal to the distance within the FMPS between the two involved modes. So clearly, in an ideal scenario we only ever perform two-mode operations between neighbours within the FMPS.

Within the 2D QRL both teleportation gadgets already utilise only nearest neighbour interactions. However, in using FMPS for simulations, modes have to be put into a 1D ordering. This is equivalent to expressing the gadgets as circuit diagrams. Thus, it is rather with respect to this 1D ordering that we want to have only nearest neighbour interactions.

\begin{figure}
    \centering
    \includegraphics[width=0.9\linewidth]{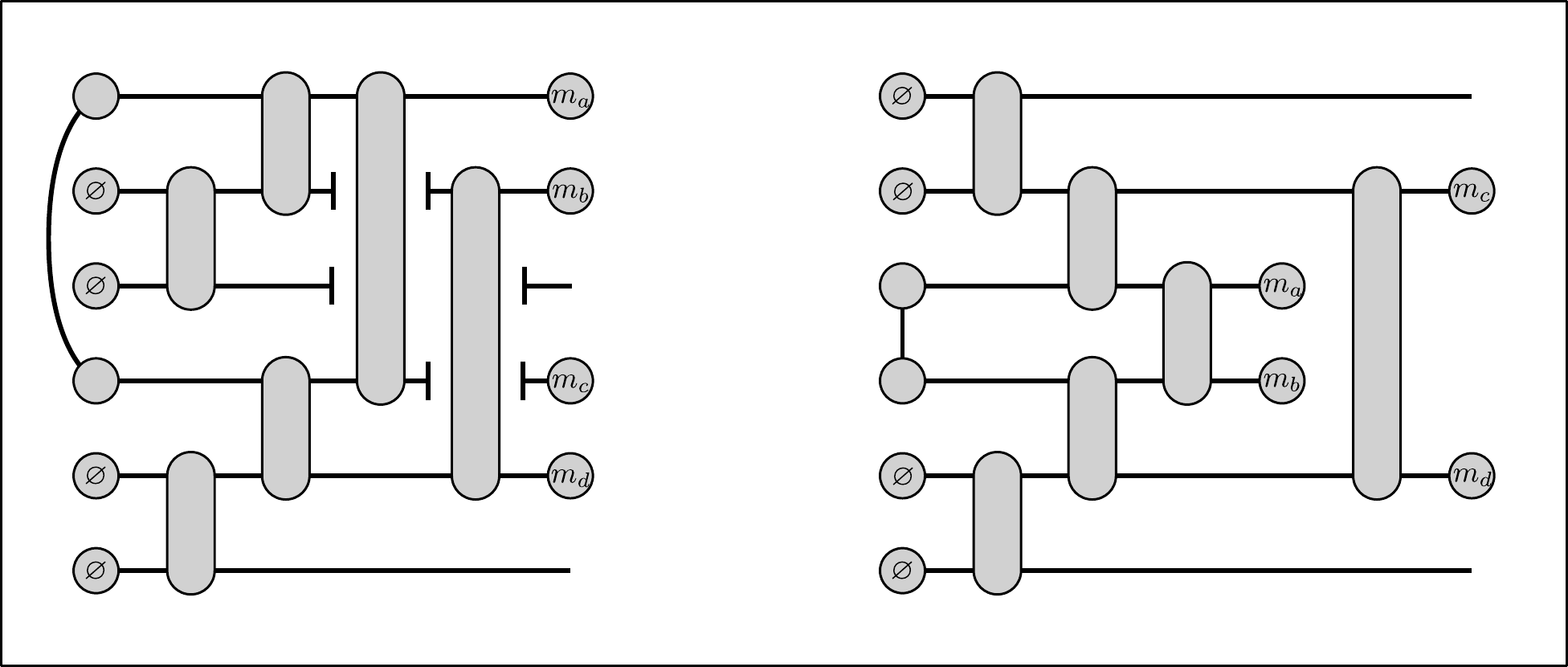}
    \caption{Two equivalent tensor network representations of the two-mode gadget. The rank-four tensors represent beam-splitters. The left network is the direct embedding of the circuit diagram presented in \cite{walshe_streamlined_2022}. The right network is a simple reordering which only requires nearest neighbour interactions.}
    \label{fig:two-mode gadget}
\end{figure}

The circuit for the single-mode gadget already has this property. The circuit for the two-mode gadget as presented in ref. \cite{walshe_streamlined_2022} does not. However, it turns out that the two-mode gadget indeed can be simulated in a satisfactory fashion. In \cref{fig:two-mode gadget} we show the tensor network diagram equivalent to the original circuit definition of the two-mode gadget, as well as an equivalent network obtained by simple reordering.

By using the reordered network for simulating the two-mode gadget, one only ever has to do two-mode operations between neighbours. The reason that the reordering works is somewhat subtle: When a mode is measured it is effectively removed from the FMPS by absorbing the result into a neighbouring mode. This means that after performing the measurements on the two middle modes 3 and 4, the initial modes 2 and 5 actually become neighbours as is also evident from the network.

\section{Fixed domain}\label{sec:domain}
For the purpose of this work, we employ a significant simplification to the general simulation methods outlined in ref. \cite{michelsen_functional_2025}. We use identical and fixed domains centred around zero for all modes throughout simulations. As long as the domain is big enough that only a negligible portion of the initial GKP qubit probability mass is lost under phase space rotation, this simplification can be taken without losing any accuracy in the simulations. The reason that the simplification is well founded is two-fold:

Firstly, as discussed in \cref{sec:decoding}, Knill error correction has the feature that the phase space envelope is directly inherited from the ancillary states being used \cite{marqversen_performance_2025}. Both teleportation gadgets, and thus all gates, in our simulations are intrinsically Knill error corrected. And since ancillary states are prepared states with consistent domain, this also applies to states at any point in between gadgets.

Secondly, since we are considering the photon damping approximation, all states have spherically symmetric phase space envelopes. As discussed in \cref{sec:noise model} this implies that the envelope is preserved under passive linear transformations. These include beam-splitters and phase rotations, which is indeed the constituents of the teleportation gadgets. Thus, states have the same fixed domain, even during the action of the gadgets.

\section{Bell-state injection}
As discussed in \cref{sec:qrl}, and shown in \cref{fig:qrl}, the QRL is really built from a lattice of Bell pairs. Since we need these Bell pair resources for every gate in our simulations, it is worth optimising the procedures involving these.

\begin{figure}
    \centering
    \includegraphics[width=0.9\linewidth]{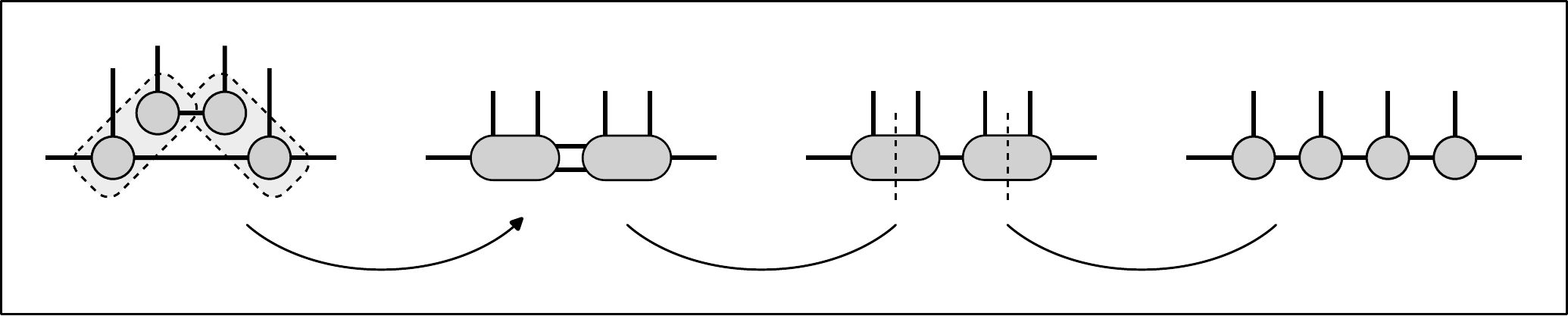}
    \caption{Tensor network representation of the procedure of inserting a two-mode MPS into a larger MPS. First the two modes are inserted into neighbouring sites. Then the two shared axes are collapsed into a single axis. Finally the two rank-four tensors are split using truncated SVD.}
    \label{fig:bell pair insertion}
\end{figure}

First consider the problem of inserting some a priori given two-mode FMPS (this will be a Bell pair) into a possibly larger FMPS. The minimum number of SVDs one can expect in general is 2. The procedure that is used in our simulations is illustrated in \cref{fig:bell pair insertion}. We note that the second step combines the two shared indices into a single one, resulting in a single index with dimension that is the product of the two. At first this might seem suboptimal and overly expensive, since in many cases this product will be quite large. However, for our purposes this turns out to never actually occur for the following reasons.

\begin{figure}
    \centering
    \includegraphics[width=0.5\linewidth]{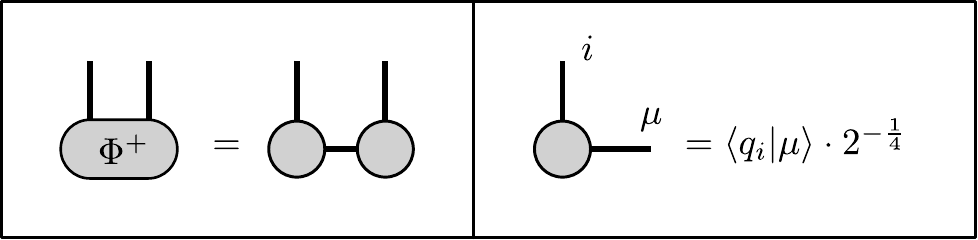}
    \caption{Two-mode MPS representation of the Bell state $\ket{\Phi^+}$. The two tensors in the right-hand side of the equation are equal and defined as shown where $\mu \in \{0, 1\}$ indexes the two logical basis states $\ket{0}$ and $\ket{1}$.}
    \label{fig:bell pair}
\end{figure}

In our case, the two-mode FMPS is a Bell state prepared by the circuit
\begin{circuit}
\begin{quantikz}[row sep={1cm, between origins}]
    \lstick{$\ket{\qunaught}$} & \bsdown{} & \rstick[2]{$\ket{\Phi^+}$} \\
    \lstick{$\ket{\qunaught}$} &           &
\end{quantikz}
\end{circuit}
The Bell states are subject to the photon dampening error model, but besides this, they are considered perfect since they are prepared states. The preparation is done by interfering two qunaught states on a beam-splitter, which is a passive linear transformation. And so we find that the prepared physical Bell state $\ket{\Phi^+_\epsilon}$ subject to photon dampening $\epsilon$ is given by
\begin{gather}
    \ket{\Phi^+_\epsilon}
        =
    BS e^{-\epsilon \hat{N}} \ket{\qunaught, \qunaught} 
        =
    e^{-\epsilon \hat{N}} BS \ket{\qunaught, \qunaught}
        =
    e^{-\epsilon \hat{N}} \ket{\Phi^+}
        \\=
    e^{-\epsilon \hat{N}} 2^{-\half} \left( \ket{00} + \ket{11} \right)
        =
    \left[ 2^{-\frac{1}{4}} e^{-\epsilon \hat{n}} \ket{0} \right]^{\tensor 2} + \left[ 2^{-\frac{1}{4}} e^{-\epsilon \hat{n}} \ket{1} \right]^{\tensor 2}
\end{gather}
This result is shown diagrammatically in \cref{fig:bell pair}. Also note that the magic Bell state $\ket{\Phi^T}$ can be obtained simply by adding the phase $e^{i \frac{\pi}{8}}$ to the second term in parentheses above.

The implications of this result are twofold: First we see that by using the construction from \cref{fig:bell pair}, the prepared Bell pair can be computed without having to apply of a beam-splitter to a two-mode state, which would come with the cost of an SVD. Secondly, we see that the Bell state has inner dimension 2, which is also the absolute minimum for any entangled state. When inserting this two-mode state into an FMPS the resulting inner dimension will thus be doubled. We conclude that if the inner dimension of the large FMPS is much smaller than half the outer dimension, then there is really no reason to reduce that dimension by doing an additional SVD.

\end{document}